\documentclass[english,floatfix,aps,unsortedaddress,superscriptaddress,twocolumn,showpacs]{revtex4}
\usepackage{longtable}
\usepackage{graphicx}

\makeatletter

\usepackage{babel}
\makeatother
\begin{document}

\title{Astrophysical factor for the neutron generator $^{13}$C($\alpha$,n)$^{16}$O reaction in the AGB stars.}

\author{E.D. Johnson}

\affiliation{Department of Physics, Florida State University, Tallahassee, FL
32306}

\author{G.V. Rogachev}

\email{grogache@fsu.edu}

\affiliation{Department of Physics, Florida State University, Tallahassee, FL
32306}

\author{A.M. Mukhamedzhanov}

\affiliation{Cyclotron Institute, Texas A\&M University, College Station, TX 77843}

\author{L.T. Baby}

\affiliation{Department of Physics, Florida State University, Tallahassee, FL
32306}

\author{S. Brown}

\affiliation{Department of Physics, Florida State University, Tallahassee, FL
32306}
\affiliation{Department of Physics, The University of Surrey, Guildford, Surrey, UK}

\author{W.T. Cluff}

\affiliation{Department of Physics, Florida State University, Tallahassee, FL
32306}

\author{A.M. Crisp}

\affiliation{Department of Physics, Florida State University, Tallahassee, FL
32306}

\author{E. Diffenderfer}

\affiliation{Department of Physics, Florida State University, Tallahassee, FL
32306}

\author{V.Z. Goldberg}

\affiliation{Cyclotron Institute, Texas A\&M University, College Station, TX 77843}

\author{B.W. Green}

\affiliation{Department of Physics, Florida State University, Tallahassee, FL
32306}

\author{T. Hinners}

\affiliation{Department of Physics, Florida State University, Tallahassee, FL
32306}

\author{C.R. Hoffman}

\affiliation{Department of Physics, Florida State University, Tallahassee, FL
32306}

\author{K.W. Kemper}

\affiliation{Department of Physics, Florida State University, Tallahassee, FL
32306}

\author{O. Momotyuk}

\affiliation{Department of Physics, Florida State University, Tallahassee, FL
32306}

\author{P. Peplowski}

\affiliation{Department of Physics, Florida State University, Tallahassee, FL
32306}

\author{A. Pipidis}

\affiliation{Department of Physics, Florida State University, Tallahassee, FL
32306}

\author{R. Reynolds}

\affiliation{Department of Physics, Florida State University, Tallahassee, FL
32306}

\author{B.T. Roeder}

\affiliation{Department of Physics, Florida State University, Tallahassee, FL
32306}

\begin{abstract}
The reaction $^{13}$C($\alpha$,n) is considered to be the main source
of neutrons for the $s$-process in AGB stars.
At low energies the cross section is dominated
by the 1/2$^{+}$ 6.356 MeV sub-threshold resonance in $^{17}$O whose contribution
is determined with a very large uncertainty of $\sim$1000\% at stellar temperatures.
In this work we performed the most precise determination of the
low-energy astrophysical $S$ factor using the indirect asymptotic normalization (ANC)
technique.
The $\alpha$-particle ANC for the sub-threshold state has been measured using the sub-Coulomb 
$\alpha$-transfer reaction ($^{6}$Li,d). Using the determined
ANC we calculated $S(0)$, which turns out to be an order of magnitude smaller 
than in the NACRE compilation.
\end{abstract}

\maketitle

About half of all elements heavier than Iron are produced in a stellar
environment through the $s$-process, which involves a series of subsequent neutron
captures and $\beta$-decays. The reaction $^{13}$C($\alpha$,n)$^{16}$O is considered to be the main source
of neutrons for the $s$-process at low temperatures in low mass stars at the Asymptotic
Giant Branch (AGB) \cite{iben76}. This is because it is exothermic and can be activated at low temperatures.
Two factors determine the efficiency of this reaction: the abundance of $^{13}$C,
and the rate of the $^{13}$C($\alpha$,n) reaction. Accurate knowledge of 
the $^{13}$C($\alpha$,n)$^{16}$O 
reaction rates at relevant temperatures (0.8 - 1.0$\times$10$^{8}$ K) would eliminate an essential uncertainty
regarding the overall neutron balance and will allow for better tests of modern Stellar models with respect
to $^{13}$C production in AGB stars (see \cite{gori01} and references therein).

The rate of the $^{13}$C($\alpha$,n) reaction at temperatures of
$\sim$10$^{8}$ K is uncertain by $\sim$300\%
\cite{angu99} due to the prohibitively small  reaction cross section at energies
below 300 keV. A directly measured $^{13}$C($\alpha$,n) cross
section is only available at energies above 279 keV (see
\cite{angu99} and references therein). Below this energy the cross
section has to be extrapolated. It was shown
\cite{drot93,angu99} that this extrapolation can be strongly
affected by the 1/2$^{+}$ sub-threshold resonance in $^{17}$O at 6.356
MeV excitation energy, which is just 3 keV below the $\alpha$
threshold. It was assumed in the recent NACRE compilation
\cite{angu99} that this resonance has a well developed $\alpha$
cluster structure. This assumption leads to a strong enhancement
of the cross section at low energies \cite{angu99}.
Recently Kubono et al.
\cite{kubo03} determined the contribution of the sub-threshold
state at $6.356$ MeV in $^{17}$O to the astrophysical factor
for the $^{13}$C($\alpha$,n) reaction at low energies by measuring the
$\alpha$-particle spectroscopic factor of this state by the
$\alpha$-transfer reaction $^{13}$C($^{6}$Li,d) at 60 MeV. The extracted spectroscopic factor
was found to be very small, $S_{\alpha} \approx 0.011$
\cite{kubo03}, making the influence of this sub-threshold state
on the astrophysical factor negligible. However, it was shown in
\cite{keel03} that the same experimental data was compatible
with a large S$_{\alpha}$ factor for the sub-threshold state in
question. It is the main goal of this work to resolve this
difference and to develop a technique which determines
the contribution of sub-threshold resonances to the ($\alpha$,n)
reaction cross sections using a model-independent approach. Until
now the ANC method has been applied to determine the astrophysical
factors for radiative capture processes \cite{mukh90,mukh97, brun99}.
Here we present the first case of application of the ANC method to
determine the astrophysical factor for the $^{13}$C($\alpha$,n)$^{16}$O reaction.

The amplitude of the reaction $x + A \to b+B$ proceeding through the 
sub-threshold resonance $F$ is given in the R-matrix approach by \cite{mukh99}
\begin{eqnarray}
M \sim \sqrt{\frac{P_{l}(k_{xA},r_{0})\,}{\mu_{xA}\,r_{0}}}\,{\tilde W}_{-\eta,l+1/2}(2\kappa _{xA}\,r_{0}) \nonumber\\
\times \frac{{\tilde C}_{xA}^{F}\,\Gamma^{1/2}_{f}(E_{bB},r_{0})}{E_{xA}+ \varepsilon + i\,\Gamma_{f}(E_{bB},r_{0})/2},
\label{resrcamplt1}
\end{eqnarray}
where $P_{l}(k_{xA},r_{0})$  is the Coulomb-centrifugal barrier
penetration factor in the entrance channel, ${\tilde
W}_{-\eta,l+1/2}(2\kappa\,r_{0}) =W_{-\eta,l +
1/2}(2\kappa\,r_{0})\,\Gamma(l + 1 + \eta)$  is the Coulomb modified
Whittaker function, $r_{0}$ the channel radius, ${\tilde
C}_{xA}^{F}=C_{xA}^F /\Gamma(l + 1 + \eta)$ stands for the Coulomb
modified ANC for the virtual decay (synthesis)  $F \leftrightarrow
x + A$, $\;\eta$ and $l$ are the Coulomb parameter and relative
orbital angular momentum of the sub-threshold bound state $(x\,A)$,
and $\Gamma_{f}(E_{bB},r_{0})$ is the resonance width for the
decay to the final channel $b+B$. We assume that the total
width of the resonance $F$ is equal to $\Gamma_{f}$, $E_{ij}=
k_{ij}^{2}/(2\,\mu_{ij})$ is the relative kinetic energy of
particles $i$ and $j$,
$\kappa= \sqrt{2\,\mu_{xA}\,\varepsilon}$ and $\varepsilon_{F}$ is
the binding energy for the virtual decay $F \to x+A$. In this case
$\Gamma_{f} \equiv\Gamma_{n}=124 \pm 12$ keV
\cite{till93} is a known neutron partial width. Thus, the ANC is the
only missing quantity needed to calculate the cross section for the
${}^{13}{\rm C}(\alpha,n){}^{16}{\rm O}$ reaction proceeding
through the sub-threshold state.
Due to the peripherality of the sub-Coulomb transfer
reactions the overall normalization of the $\alpha$-transfer
reaction cross section is determined by the product of the squares
of the initial and final ANCs rather than the spectroscopic
factors. The initial ANC for the $\alpha + d \to {}^{6}{\rm Li}$
is known, $(C^{{}^{6}{\rm Li}}_{\alpha\,d})^2 = 5.3 \pm
0.5$ fm$^{-1}$ \cite{blok93}. Hence, by normalizing the DWBA
cross section to the experimental one we can determine the ANC for
$\alpha + {}^{13}{\rm C} \to {}^{17}{\rm O}$ (6.356 MeV, 1/2$^+$).

\begin{figure}[top]
\includegraphics[width=0.9\columnwidth]{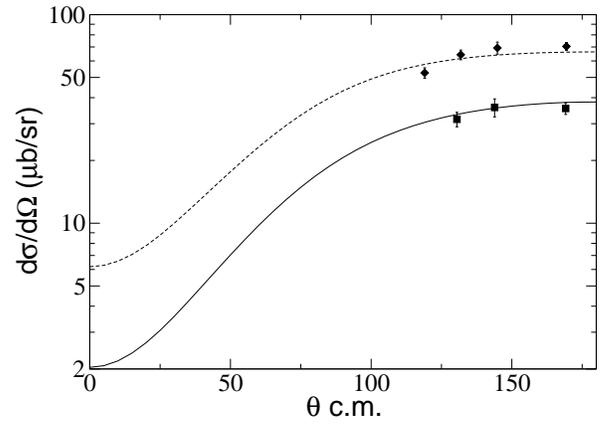}
\caption{\label{fig:adist} Angular distribution of the $^{13}{\rm C}(^{6}{\rm Li},d){}^{17}{\rm O}(1/2^{+})$ reaction.
Data taken at 8.5 MeV of $^{13}$C are shown
as diamonds, and 8.0 MeV data as boxes. Dashed and solid lines are
DWBA calculations at $^{13}$C energies 8.31 and 7.81 MeV (see text).}
\end{figure}
\begin{figure}[top]
\includegraphics[%
  width=0.9\columnwidth]{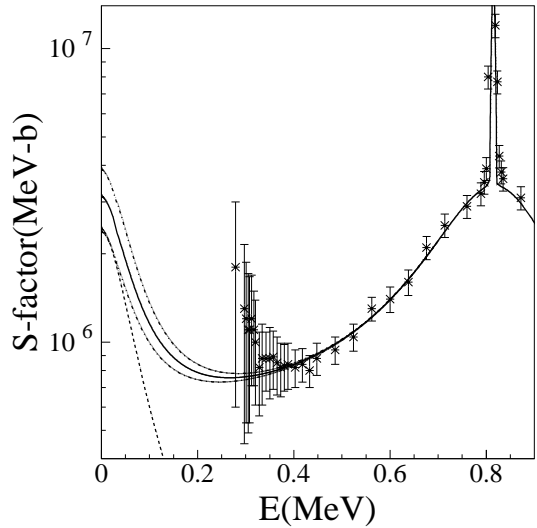}
\caption{\label{fig:sfactor} S-factor of the $^{13}$C($\alpha$,n) reaction.
The experimental data from direct measurements, corrected for electron
screening, are from \cite{angu99}. The contribution of 1/2$^{+}$ state is shown
as the dashed curve. The dash-dotted curves represent 26\% uncertainty band.}
\end{figure}

In this Letter, we report the application of the
ANC technique to determine the astrophysical $S$ factor of the $^{13}{\rm C}(\alpha,n){}^{16}{\rm O}$
reaction at astrophysically relevant energies by measuring the ANC for the virtual synthesis 
$\alpha + {}^{13}{\rm C} \to {}^{17}{\rm O}$ (6.356 MeV, 1/2$^{+}$) using the
$\alpha$-transfer reaction $^{6}$Li($^{13}$C,d), performed at two sub-Coulomb energies, 8.0 and 8.5 MeV, of $^{13}$C
at the Florida State University Tandem-LINAC facility. The choice of inverse kinematics, $^{13}$C beam and $^{6}$Li target, 
allowed measurements to be made at very low energies in the c.m.,
and to avoid background associated with the admixture of $^{12}$C
in the $^{13}$C target.
\begin{figure}[top]
\includegraphics[width=0.9\columnwidth,keepaspectratio]{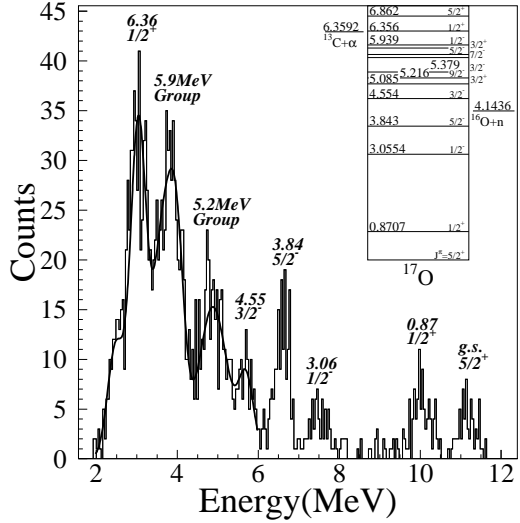}
\caption{\label{fig:spectrum} Spectrum of deuterons from the $^{6}$Li($^{13}$C,d)
reaction, at 6$^{\circ}$ with the 8.5 MeV $^{13}$C beam. The inset shows the 
level scheme of $^{17}$O \protect\cite{till93}.
The solid line is a Gaussian fit.}
\end{figure}
Angular distributions of the deuterons from transfer reactions
at sub-Coulomb energies in inverse kinematics peak at forward angles.
Four Si $\Delta E-E$ telescopes were positioned at forward angles
to identify deuterons. Thicknesses of the $\Delta E$ detectors were in
the range from 15 to 25 $\mu$m. 50 $\mu g/cm^{2}$ Li targets
(enriched to 98 \% of $^{6}$Li) were prepared and transported into
a scattering chamber under vacuum to prevent oxidation. The
telescope at the smallest angle (6$^{\circ}$ in Lab. frame) was
shielded from the Rutherford scattering of $^{13}$C on Li target with a 5
$\mu$m Havar foil.
A spectrum of deuterons at 6$^{\circ}$ (which corresponds to 169$^{\circ}$
in the c.m. for the 1/2$^{+}$ 6.356 MeV state) at a beam energy 8.5 MeV is shown in Fig. \ref{fig:spectrum}.
The typical experimental resolution in the c.m. system (mainly
defined by the 380 keV energy loss of the $^{13}$C beam in the $^{6}$Li
target) was about 250 keV (FWHM).
The angular distributions of 
$^{6}$Li($^{13}$C,d)$^{17}$O(1/2$^{+}$, 6.356 MeV) are shown
in Fig. \ref{fig:adist}. Their shape is
typical for sub-Coulomb transfer reactions. Absolute
normalization of the cross section was performed by measuring the
elastic scattering of 6.868 MeV protons on the $^{6}$Li target. The cross
section of this reaction at 95$^{\circ}$ is known with 3\% accuracy
\cite{Bing70}. Each telescope was sequentially placed at 95$^{\circ}$
and the product of the target thickness times the telescope solid angle
($t\times\Delta\Omega$) was determined for each telescope.
The code FRESCO (version FRXY.3h) \cite{thom88} was used to calculate the angular distribution 
of the $^{13}$C($^{6}$Li,d) reaction in the Distorted Wave 
Born Approximation (DWBA) approach.
The DWBA calculations were performed for beam energies at the
center of the target, 8.31 MeV for 8.50 MeV measurements and 7.81
MeV for 8.0 MeV energy of the beam.
It was found that normalization factor 
is the same for both energies, indicating that Compound Nucleus 
mechanism plays only minor (if any) role.
The extracted ANC, unlike the
spectroscopic factor, does not depend on the number of nodes
of the ${\alpha-{}^{13}{\rm C}}_{1/2^{+}}$ bound state wave
function, or the geometrical parameters of the Woods-Saxon potential.
Parameters of optical model potentials of the usual Woods-Saxon form, used in the DWBA calculations, are given
in Table \ref{tab:potentials}. The LC1 potential was used for the $^{6}$Li+$^{13}$C
channel. It reproduces experimental data on elastic scattering of
$^{6}$Li by $^{13}$C at energies ranging from 3 to 23 MeV
in c.m. Experimental data on elastic scattering of deuterons on
$^{17}$O at low energies are not available. Thus, several potentials for the
d +$^{17}$O channel were used \cite{Drai75,Li76}.
Angular distributions shown in Fig. \ref{fig:adist} were calculated with the DO1 potential,
however it was verified that other potentials \cite{Drai75,Li76} produce essentially identical results,
with variations in normalization factor of less than 7\%. This demonstrates
that the transfer reaction cross section at sub-Coulomb energy only weakly depends on
the parameters of the optical potentials, which was the main point of
making this measurement. In fact, calculation with no nuclear part
of the optical potentials changes the absolute value of the cross section
at large angles by only $\sim$40\%. Further investigation of the cross section sensitivity 
to the parameters of the optical potentials was performed and was found to be
less than 20\% if parameters kept within reasonable limits. We found no sensitivity of the extracted ANC to
the parameters of the core-core DC1 interaction potential in the full DWBA transition operator.
\begin{table*}[tbh]
\caption{\label{tab:potentials}Parameters of Optical potentials used in DWBA
calculations.}
\begin{tabular}{cccccccccccccc}
\hline
\hline
        &         &  $V_{0}$   &      $a_{V}$&    $r_{V}$&    $W$&     $W_{S}$&    $a_{W}$&    
	          $r_{W}$&    $r_{c}$ &  $V_{so}$&   $a_{so}$&   $r_{so}$ \\
Channel & Potential&  (MeV)& (fm)&  (fm)&  (MeV)&  (MeV)&  (fm)&  (fm)&  (fm)&  (MeV)&  (fm)&  (fm)& Ref.\\
\hline
$^{6}$Li +$^{13}$C& LC1& 134.0& 0.68& 1.50& -& 11.1& 0.68& 1.50& 1.50& -& -& -& \cite{Poli76}  \\
d +$^{17}$O& DO1& 105.0& 0.86& 1.02& -& 15.0& 0.65& 1.42& 1.40& 6.0& 0.86& 1.02& \cite{Drai75} \\
d + $^{13}$C& DC1& 79.5& 0.80& 1.25& 10.0& -& 0.80& 1.25& 1.25& 6.0& 0.80& 1.25& \cite{Putt71} \\
\end{tabular}
\newline
$R=r_{0}A_{T}^{\frac{1}{3}}$; $r_{0}=1.25$ fm and $a=0.68$ fm were used for $\alpha+d$ and $\alpha+ {}^{13}$C
form factor potentials with $V$ fitted to reproduce binding energy.
\end{table*}
Our determined Coulomb-modified ANC squared for  
$^{13}{\rm C} + \alpha \to {}^{17}{\rm O}(1/2^{+}$,6.356 MeV) is
$(\tilde C_{\alpha\,{}^{13}{\rm C}}^{{}^{17}{\rm O}(1/2^{+})})^{2}=0.89\pm 0.23 $ fm${}^{-1}$. 
The contribution of the 1/2$^{+}$ state to the astrophysical $S$ factor 
calculated using Eq. (\ref{resrcamplt1}) is shown as a dashed curve in Figure \ref{fig:sfactor}.
It was verified that this result is insensitive to variations of the channel radius.

Five sources of uncertainty associated with the S(0) factor of the 6.356 MeV 1/2$^{+}$
state (same as for the ANC), can be identified: 7\% statistical uncertainty, 7\% combined systematical uncertainty
in determination of the $t\times\Delta\Omega$
value (target thickness times solid angle, as described above),
20\% uncertainty associated with theoretical analysis 
(uncertainty of the square of the ANC due to the variation of the optical potential parameters), 10\%
uncertainty in the total resonance width and 10\% uncertainty due to the initial $\alpha$-d ANC. 
Thus, the S(0) value of 1/2$^{+}$ resonance determined in this
experiment is
$(2.5\pm0.7)\times10^{6}$
MeV$\times$b. The S(0) of the 6.356 MeV 1/2$^+$ state in $^{17}$O determined in this experiment is
ten times smaller than that adopted in
the NACRE compilation \cite{angu99} and a factor of 5 larger than in \cite{kubo03}.
\begin{figure}[top]
\includegraphics[%
  width=0.9\columnwidth,
  keepaspectratio]{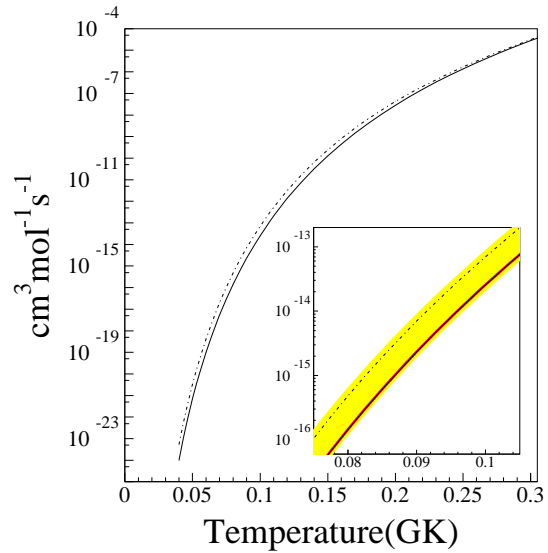}
\caption{\label{fig:rate}Rate of the $^{13}$C($\alpha$,n) reaction. The dash-dotted curve is from
\cite{angu99}, solid curve is the rate obtained in this work. Yellow band in inset represents
uncertainties from \cite{angu99}, new uncertainties are shown in red.}
\end{figure}
The two channel R-matrix approach was used to calculate the total S-factor
of the $^{13}$C($\alpha$,n) process. All resonances from 4.14 MeV (neutron
decay threshold) to 8.2 MeV excitation energy in $^{17}$O were included.
Parameters of the resonances were taken from \cite{till93}. The $\alpha$-reduced
widths of resonances were fitted to reproduce the $^{13}$C($\alpha$,n)
experimental data \cite{drot93}. It was found that it is necessary to introduce
a constant $0.4\times10^{6}$ MeV$\times$b non-resonance contribution
to fit the experimental $S$ factor data at the lowest energy region.
The solid curve in Fig. \ref{fig:sfactor} corresponds to the best R-matrix
fit, obtained as explained above, with contribution from $1/2^{+}$
state.
Our calculated reaction rates are shown in Figure \ref{fig:rate}.
The best fit rate is shown
as a solid curve in Fig. \ref{fig:rate} in comparison with the NACRE adopted
rate (dash-dotted curve). At temperatures above 0.3 GK, where contribution
of 1/2$^{+}$ is small, it is identical to the curve adopted in NACRE,
however, at temperatures which are most significant for the $s$-process
in AGB stars, 0.08-0.1 GK, the reaction rate is smaller by a factor
of 3 than that adopted in the NACRE compilation. Uncertainty in this astrophysically
important reaction rate is now reduced from $\sim$300\%
to 15\%. The inset in Fig.
\ref{fig:rate} shows the NACRE compilation adopted rate (dash-dotted
curve) with uncertainty band (yellow) and the reaction rate obtained in
this work (solid line) with uncertainty band shown in red. Numerical
values of reaction rate for 0.08 - 0.1 GK temperature range are given
in Table \ref{tab:rate}.
\begin{table}[tbh]
\caption{\label{tab:rate}The rate of $^{13}$C($\alpha,n)$ reaction at temperatures
from 0.08 to 0.1 GK. The rate obtained in this work is compared with
the rate published in the NACRE \cite{angu99} compilation. Units are {[}cm$^{3}$mol$^{-1}$s$^{-1}${]},
$\exp$ stands for 10$^{exp}$. High and low values were calculated
assuming $26$\% uncertainty of 1/2$^{+}$ 6.356 MeV resonance contribution.}
\begin{tabular}{c|ccc|ccc|c}
\hline
\hline
       &     &   This&   work&      &  NACRE&      &     \\
T$_{9}$&  low&  adopt&   high&   low&  adopt& high & exp \\
\hline
0.08   & 1.34&   1.44&   1.56&  1.22&   4.80& 5.80 & -16 \\
0.09   & 2.18&   2.32&   2.50&  2.03&   6.99& 8.45 & -15 \\
0.10   & 2.42&  2.56 &   2.73&  2.28&   6.99& 8.49 & -14 \\
\end{tabular}
\end{table}

In summary, in this Letter we developed an indirect technique which allows
measurement of the astrophysical S(0) factor of sub-threshold, particle
unbound resonances and applied this technique to measure the contribution
of the $1/2^{+}$ 6.356 MeV resonance in $^{17}$O to the $^{13}$C($\alpha$,n)
reaction rate at stellar temperatures. Combination of the sub-Coulomb
$\alpha$-transfer reaction and application of the ANC technique in the
analysis of experimental data practically eliminates all dependence of
the results on model parameters, making this approach a very
valuable tool for future studies of astrophysically important reaction
rates with both stable and radioactive beams. The $^{13}$C($\alpha$,n)
reaction rate at stellar temperatures was found to be lower by a factor
of 3 than previously adopted \cite{angu99}, also uncertainty in this reaction
rate was greatly reduced. It would be of great interest to incorporate
the new reaction rate into the $s$-process calculations in
AGB stars.

Authors are grateful to Profs. I. Thompson and M. Wiescher for valuable discussions
and acknowledge the financial support provided by NSF under grant PHY-04-56463,
and by the U. S. Department of Energy under Grant No. DE-FG02-93ER40773.

\bibliography{myrefs}

\end{document}